\documentclass[graybox]{svmult}

\usepackage{mathptmx}       % selects Times Roman as basic font
\usepackage{helvet}         % selects Helvetica as sans-serif font
\usepackage{courier}        % selects Courier as typewriter font
\usepackage{type1cm}        % activate if the above 3 fonts are
\usepackage{makeidx}         % allows index generation
\usepackage{graphicx}        % standard LaTeX graphics tool
\usepackage{multicol}        % used for the two-column index
\usepackage[bottom]{footmisc}% places footnotes at page bottom
\usepackage{aas_macros}
\usepackage{natbib}
\usepackage{amsmath}
\usepackage{amssymb}
\usepackage{subfigure}

\makeindex             % used for the subject index
\begin{document}

\title*{Molecular Gas in the Outskirts}
\author{Linda C.\ Watson and Jin Koda}
\institute{Linda C.\ Watson \at European Southern Observatory, Alonso de C\'ordova 3107, Vitacura, Casilla 19001, Santiago, Chile; \\
\email{lwatson@eso.org}
\and Jin Koda \at NAOJ Chile Observatory, Joaqu\'in Montero 3000 Oficina 702, Vitacura, Santiago, Chile; \\
Joint ALMA Office, Alonso de C\'ordova 3107, Vitacura, Casilla 19001, Santiago, Chile; \\ 
Department of Physics and Astronomy, Stony Brook University, Stony Brook, NY 11794, USA; \\ 
\email{jin.koda@stonybrook.edu}} 
\maketitle

% Abstract
% ======
\abstract{The outskirts of galaxies offer extreme
environments where we can test our understanding of the formation,
evolution, and destruction of molecules and their relationship with
star formation and galaxy evolution. We review the basic equations
that are used in normal environments to estimate physical 
parameters like the molecular gas mass from CO line emission and
dust continuum emission. Then we discuss how those estimates may be
affected when applied to the outskirts, where the 
average gas density, metallicity, stellar radiation field, and
temperature may be lower. We focus on observations of molecular gas in the
outskirts of the Milky Way, extragalactic disk galaxies, early-type
galaxies, groups, and clusters. The scientific results show the
versatility of molecular gas, as it has been used to trace Milky Way
spiral arms out to a galactocentric radius of 15\,kpc, to study star
formation in extended ultraviolet disk galaxies, to probe galaxy
interactions in polar ring S0 galaxies, and to investigate ram pressure
stripping in clusters. Throughout the Chapter, we highlight the
physical stimuli that accelerate the formation of molecular gas,
including internal processes such as spiral arm compression and
external processes such as interactions.}

% Body
% ====

% \section{Section Heading}
% \subsection{Subsection Heading}
% \paragraph{Paragraph Heading}
% \subparagraph{Subparagraph Heading} In order to avoid ...

\section{Introduction}

Despite early discoveries of OB stars and molecular gas in the
outer Milky Way (MW; e.g., \citealt{fich84, brand88}), not much attention
had been paid to molecular gas 
in galaxy outskirts primarily because there was a notion that
virtually no star formation occurs there. This notion was altered
entirely by the {\it Galaxy Evolution Explorer} ({\it GALEX})\index{{\it GALEX}}, which
revealed that ultraviolet emission often extends far beyond the edges of
optical disks (namely, extended ultraviolet disks, or XUV
disks\index{XUV disk}; \citealt{Thilker:2005ff, Gil-de-Paz:2007eu}). The UV
emission suggests the presence of massive stars, at least B stars, and
hence that there was recent star formation within the lifetime of B stars
($\sim 100$\,Myr). These young stars must have been born nearby,
perhaps requiring unnoticed molecular gas and clouds somewhere in the
extended galaxy outskirts. Average gas densities there are extremely
low compared to typical star-forming regions within the MW. Understanding the
conditions of parental molecular gas in such an extreme condition 
is vital to expand our knowledge of the physics of star formation.
We need to understand the internal properties of molecular clouds,
including the atomic-to-molecular gas phase transition, the
distribution of molecular clouds, and the external environment in
galaxy outskirts. 

A blind search for molecular gas has been difficult for
the large outskirts of nearby galaxies due to the limited capability of existing facilities.
The Atacama Large Millimeter/submillimeter Array (ALMA) improved the sensitivity remarkably,
but even ALMA would need to invest hours to days to carry out a large areal search for molecular gas over extended disks.
This review summarizes the current knowledge on molecular gas and star formation
in the outskirts, but this research field is still in a phase of discovery.
The space to explore is large, and more systematic understanding will become possible
with future observations.

Studies of molecular gas in the outskirts will also reveal the yet
unknown physical properties of the interstellar medium (ISM) in the
outskirts. Most observational tools were developed and calibrated in
the inner parts of galactic disks and may not be applicable as they
are to the outskirts. Many studies are subject to {\it
systematic biases}, especially when molecular gas in the outskirts
is compared with inner disks. For example, the rotational
transition of carbon monoxide (CO) is often used to measure the mass
of molecular gas in normal galaxies; however, its presence and
excitation conditions depend on the metal abundance, stellar radiation
field, internal volume and column densities, and kinetic temperature,
all of which may change in the outskirts.

In this review, we start from a summary of how the ISM evolves in the
inner parts of the MW and nearby galaxies with an emphasis on
molecular gas (Sect.~\ref{sec:inout}). We then discuss the observational methods, including
the equations needed to plan for a future observational search of
molecular gas with a radio telescope (Sect.~\ref{sec:clouds}). We explain the 
potential effects of applying these equations under the extreme 
conditions in galaxy outskirts, which may cause systematic biases when the ISM is
compared between galaxies' inner parts and outskirts (Sect.~\ref{sec:ISMextreme}).
Although not many observations have been carried out in galaxy outskirts,
we summarize the current state of molecular gas observations in spiral (Sect.~\ref{sec:disks})
and elliptical galaxies (Sect.~\ref{sec:ellipticals}) and in galaxy groups and clusters (Sect.~\ref{sec:groups}).
We finish the review with possible future directions
(Sect.~\ref{sec:future}). The term ``outskirts" is abstract and has
been used differently in different contexts. In this review we use
this term for the area beyond the optical radius of galaxy, e.g.,
beyond $r_{25}$, which is the radius where the $B$-band surface
brightness of a galaxy falls to $25\,\rm mag\, arcsec^{-2}$. We should,
however, note that in some circumstances $r_{25}$ is not defined well,
and we have to rely on a loose definition of ``outskirts".

The measurements of gas properties, such as molecular mass, often depend on some
assumptions of the gas properties themselves.
However, galaxy outskirts are an extreme environment, and the
assumptions based on previous measurements in inner disks may not be appropriate.
This problem needs to be resolved iteratively by adjusting the
assumptions to match future observations. We therefore spend a 
number of pages on the methods of basic measurements
(Sect.~\ref{sec:clouds}), so that the equations and assumptions can be
revisited easily in future studies. Readers who already understand the
basic methods and assumptions may skip Sect.~\ref{sec:clouds}
entirely and move from Sect.~\ref{sec:inout} to Sect.\ref{sec:disks}.

\section{Molecular Gas from the Inner to the Outer Regions of Galaxies}\label{sec:inout}

The most abundant molecule H$_2$\index{H$_{2}$ molecule} does not have
significant emission at the cold temperatures that are typical in
molecular clouds ($<30$\,K).  Hence, the emission from CO\index{CO molecule}, the
second-most abundant molecule, is commonly used to trace molecular gas.
Molecular gas is typically concentrated toward the centres of galaxies and
its surface density decreases with galactic radius (\citealt{Young:1991aa, Wong:2002lr}).
The gas phase changes from mostly molecular in the central regions
to more atomic in the outer regions (\citealt{Sofue:1995fk, Koda:2016aa, Sofue:2016aa}).
These trends apparently continue into the outskirts, as H{\sc i} disks often
extend beyond the edges of optical disks (\citealt{Bosma:1981aa}).

We may infer the properties of gas in the outskirts by extending our knowledge from
the inner disks.
Recently, \citet{Koda:2016aa} concluded that the H{\sc i}-H$_2$ gas phase
transition\index{H{\sc i}-H$_2$ gas phase transition} between spiral arm and
interarm regions changes as a function of radius in the MW 
and other nearby galaxies. In the molecule-dominant inner parts, the
gas remains highly molecular as it moves from an interarm region into
a spiral arm and back into the next interarm region. Stellar feedback
does not dissociate molecules much, and perhaps the coagulation and
fragmentation of molecular clouds dominate the evolution of the ISM at
these radii. The trend differs in the outer regions where the gas
phase is atomic on average. The H{\sc i} gas is converted to H$_2$ in spiral
arm compression and goes back into the H{\sc i} phase after passing spiral
arms. These different regimes of ISM evolution are also seen in the
LMC, M33, and M51, depending on the dominant gas phase there 
(\citealt{Heyer:1998kx, Engargiola:2003jo, Koda:2009wd, Fukui:2009lr, Tosaki:2011fk, Colombo:2014uq}).

Even in regions of relatively low gas densities, a natural fluctuation
may occasionally lead to gravitational collapse into molecular gas and clouds.
For example, many low-density dwarf galaxies show some molecular gas and star formation.
However, some stimulus, such as spiral arm compression, seems necessary to accelerate
the H{\sc i} to H$_2$ phase transition. In addition to such internal stimuli, there are
external stimuli, such as interactions with satellite galaxies, 
which may also trigger the phase transition into molecular gas in the outskirts.

\section{Molecular ISM Masses: Basic Equations} \label{sec:clouds}

The molecular ISM is typically cold and is observed at radio wavelengths.
To search for the molecular ISM in galaxy outskirts one needs to be familiar
with conventional notations in radio astronomy.
Here we summarize the basic equations and assumptions that 
have been used in studies of the molecular ISM in traditional environments, such as
in the MW's inner disk. In particular, we focus on the $J=1-0, 2-1$ rotational transitions
of CO molecules and dust continuum emission at
millimetre/sub-millimetre wavelengths.
The molecular ISM in galaxy outskirts may have different properties from those in the inner disks.
We discuss how expected differences could affect the measurements with
CO $J=1-0, 2-1$, and dust continuum emission.

%%%

\subsection{Brightness Temperature, Flux Density and Luminosity}

\begin{figure}[h]
\begin{center}
\includegraphics[scale=.7]{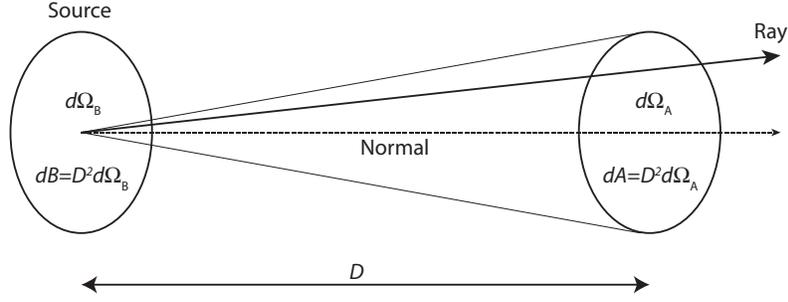}
\end{center}
\caption{Definitions of parameters. The rays emitted from the source with the area $dB = D^2 d\Omega_{\rm B}$
pass through the solid angle $d\Omega_{\rm A}$ (or the area $D^2 d\Omega_{\rm A}$) at the distance of $D$}
\label{fig:rad} 
\end{figure}

The definitions of brightness temperature $T_{\nu}$\index{brightness
temperature}, brightness $I_{\nu}$\index{brightness}, flux density
$S_{\nu}$\index{flux density}, and luminosity
$L_{\nu}$\index{luminosity} are often confusing. 
%To understand their relations
It is useful to go back to the amount of energy ($dE$) that passes through an aperture (e.g., detector, or sometimes the $4\pi$ sky area),
\begin{equation}
dE = I_{\nu} d\Omega_{\rm B} dA dt d\nu = \left\{ \left[ I_{\nu} d\Omega_{\rm B} \right] dA \right\} dt d\nu = \left\{ S_{\nu} dA \right\} dt d\nu = L_{\nu} dt d\nu, \label{eq:energy}
\end{equation}
where $S_{\nu} = \int I_{\nu} d\Omega_{\rm B}$ and $L_{\nu} = \int \int I_{\nu} d\Omega_{\rm B} dA$ (see Fig.~\ref{fig:rad}). 
The $dt$ and $d\nu$ denote unit time and frequency, respectively.
The $d\Omega_{\rm B}$ is the solid angle of the source and has the relation with the physical area $dB=D^2 d\Omega_{\rm B}$
with the distance $D$.
Similarly, $dA=D^2 d\Omega_{\rm A}$ using the solid angle of the aperture area seen from the source $d\Omega_{\rm A}$.
The aperture $dA$ can be a portion of the $4\pi$ sky sphere as it is seen from the source and is $4\pi D^2$
when integrated over the entire sphere to calculate luminosity.
The $dA$ could also represent an area of a detector (or a pixel of a detector).
%The black body radiation $B_{\nu}(T)$ is an example of brightness (i.e., $B_{\nu}(T) \in I_{\nu}$).

The flux density $S_{\nu}$ is often expressed in the unit of ``Jansky (Jy)'', which is equivalent to
``$10^{-23}\,\rm erg\, s^{-1}\,cm^{-2}\,Hz^{-1}$".
An integration of $I_{\nu}$ over a solid angle $d\Omega_{\rm B}$
(e.g., telescope beam area or synthesized beam area)
provides $S_{\nu}$. In reverse, $I_{\nu}$ is $S_{\nu}$ divided by the solid angle $\Omega_{\rm B}$ [$= \int d\Omega_{\rm B}$].
Therefore, the brightness $I_{\nu}$ [$= S_{\nu}/ \Omega_{\rm B}$] is expressed in the unit of ``Jy/beam".

%\paragraph{Brightness Temperature}

The brightness temperature $T_{\nu}$ is the temperature that makes the black body function
$B_{\nu}(T_{\nu})$ have the same brightness as the observed $I_{\nu}$
at a frequency $\nu$ (i.e., $I_{\nu}=B_{\nu}(T_{\nu})$), even when
$I_{\nu}$ does not follow the black body law! In the Rayleigh-Jeans
regime  ($h\nu \ll kT$),
\begin{equation}
T_{\nu}= \frac{c^2 }{2 \nu^2 k} I_{\nu}= \frac{c^2 }{2 \nu^2 k} \left( \frac{S_{\nu}}{\Omega_{\rm B}} \right). \label{eq:t_s}
\end{equation}
The $T_{\nu}$ characterizes radiation and is {\it not necessarily} a physical temperature of an emitting body.
However, if the emitting body is an optically thick black body and is filling the beam $\Omega_{\rm B}$,
$T_{\nu}$ is equivalent to the physical temperature of the emitting
body when the Rayleigh-Jeans criterion is satisfied.

The $T_{\nu}$ is measured in ``Kelvin". This unit is convenient in radio astronomy since radio single-dish observations
calibrate a flux scale in the Kelvin unit using hot and cold loads of known temperatures.
Giant molecular clouds (GMCs) in the MW have a typical temperature of
$\sim$10\,K (\citealt{Scoville:1987vo}), and the black body radiation $B_{\nu}(T)$ at this temperature
peaks at $\nu \sim 588\,\rm GHz$ ($\sim 510\mu\rm m$).
Therefore, most radio observations of molecular gas are in the Rayleigh-Jeans range.

%\paragraph{Numerical Expressions}

A numerical expression of Eq.~(\ref{eq:t_s}) is useful in practice,
\begin{equation}
\left( \frac{T_{\nu}}{\rm K} \right)=  13.6 \left( \frac{\lambda}{\rm mm} \right)^2  \left( \frac{S_{\nu}}{\rm Jy} \right) \left( \frac{b_{\rm maj} \times b_{\rm min}}{\rm 1" \times 1"} \right)^{-1}.\label{eq:numRJ}
\end{equation}
The last term corresponds to $\Omega_{\rm B}$ in Eq.~(\ref{eq:t_s}) and is calculated as
\begin{equation}
\Omega_{\rm B} = \frac{\pi b_{\rm maj} b_{\rm min}}{4\ln 2} \sim 1.133 b_{\rm maj} b_{\rm min},
\end{equation}
which represents the area of interest (e.g., source size, telescope beam) as a 2-d Gaussian with the major and minor axis FWHM diameters of $b_{\rm maj}$ and $b_{\rm min}$, respectively.
Equation~(\ref{eq:numRJ}) is sometimes written with brightness as
\begin{equation}
\left( \frac{T_{\nu}}{\rm K} \right)=  13.6 \left( \frac{\lambda}{\rm mm} \right)^2  \left( \frac{I_{\nu}}{\rm Jy/beam} \right) \left( \frac{b_{\rm maj} \times b_{\rm min}}{\rm 1" \times 1"} \right)^{-1},
\end{equation}
where in this case the last term is for the unit conversion from
``beam" into arcsec$^2$, and $b_{\rm maj}$ and $b_{\rm min}$ must refer
to the telescope beam or synthesized beam.

%%%
\subsection{Observations of the Molecular ISM using CO Line Emission}

Molecular hydrogen (H$_2$) is the principal component of the ISM at a high density, $>100\rm\, cm^{-3}$.
This molecule has virtually no emission at cold temperatures.
Hence, CO emission is typically used to trace the molecular ISM.
Conventionally, the molecular ISM mass $M_{\rm mol}$\index{molecular
mass} includes the masses of helium and other elements. $M_{\rm
mol}=1.36\,M_{\rm H_2}$ is used to convert the H$_2$ mass into $M_{\rm
mol}$.

\subsubsection{CO($J=1-0$) Line Emission}\index{CO($J=1-0$) line emission}

The fundamental CO rotational transition $J=1-0$ at $\nu_{\rm CO}(1-0)=115.271208$\,GHz
has been used to measure the molecular ISM mass since the 1980s.
For simplicity we omit ``CO($1-0$)" in subscript and instead write ``10". Hence, $\nu_{\rm CO}($1-0$)=\nu_{10}$.

The dynamical masses of GMCs and their CO($1-0$) luminosities are linearly correlated
in the MW's inner disk (\citealt{Scoville:1987lp, solomon87}).
If a great majority of molecules reside in GMCs, the CO($1-0$) luminosity $L_{\rm 10}^{\prime}$
integrated over an area (i.e., an ensemble of GMCs in the area) can be linearly translated to the molecular mass $M_{\rm mol}$,
\begin{equation}
M_{\rm mol}=\alpha_{\rm 10} L_{\rm 10}^{\prime},
\end{equation}
where $\alpha_{10}$ (or $X_{\rm CO}$\index{$X_{\rm CO}$}; see below)
is a mass-to-light ratio and is called the CO-to-H$_2$ conversion
factor\index{conversion factor} (\citealt{Bolatto:2013ys}).

By convention we define $L_{10}^{\prime}$, instead of $L_{10}$ (Eq.\ref{eq:energy}).
With the CO($1-0$) brightness temperature $T_{10}$ (instead of $I_{\nu}$ or $I_{10}$),
velocity width $d v$ (instead of frequency width $d\nu$), and beam area in physical scale $ dB= D^2 d\Omega_{\rm B}$,
it is defined as
\begin{equation}
L_{10}^{\prime} \equiv \int \int T_{10}dv dB= \frac{c^2 }{2 \nu_{10}^2 k} \left[ \int S_{10} dv \right] D^2 ,\label{eq:lum_s10}
\end{equation}
where we used Eq.~(\ref{eq:t_s}) for $T_{10}$.
The molecular mass is
\begin{equation}
M_{\rm mol} = \alpha_{\rm 10} \frac{c^2}{2 \nu_{10}^2 k }  \left[ \int S_{10} dv \right]  D^2. \label{eq:mass_s10}
\end{equation}

%\paragraph{Numerical Expressions}
Numerically, this can be expressed as
\begin{equation}
\left( \frac{M_{\rm mol}}{M_{\odot}} \right) = 1.1 \times 10^{4} \left( \frac{\alpha_{\rm 10}}{4.3\,M_{\odot}\,{\rm pc^{-2} [K\cdot km/s]}^{-1}} \right)\left( \frac{\int S_{10}dv}{\rm Jy \cdot km/s} \right) \left( \frac{D}{\rm Mpc}\right)^2.\label{eq:m_s10_num}
\end{equation}
Note that $S_{10}$ [$=\int I_{10}d\Omega_{\rm B}$] is an integration over an area of interest (or summation over all pixels within the area).
The $\alpha_{\rm 10}=4.3\,M_{\odot}\,{\rm pc}^{-2}$ corresponds to the conversion factor of
$X_{\rm CO}=2.0\times 10^{20}\rm \, cm^{-2}\,[K\cdot km/s]^{-1}$ multiplied by the factor of 1.36 to account for
the masses of helium and other elements. 
$\alpha_{\rm 10}$ includes helium, while $X_{\rm CO}$ does not.
The calibration of $\alpha_{\rm 10}$ (or $X_{\rm CO}$) is discussed in \citet{Bolatto:2013ys}.

A typical GMC in the MW has a mass of $4\times10^5\,M_{\odot}$ and $dv=8.9\,\rm km/s$ (FWHM)
(\citealt{Scoville:1987vo}), which is $\int S_{10}dv\sim1.5\,\rm Jy\,km/s$ or $S_{10}\sim 170\,\rm mJy$ at $D=5\,\rm Mpc$.

\subsubsection{CO($J=2-1$) Line Emission}\index{CO($J=2-1$) line emission}

The CO($J=2-1$) emission (230.538\,GHz) is also useful for a rough estimation of molecular mass
though an excitation condition may play a role (see below).
We can redefine Eq.~(\ref{eq:mass_s10}) for CO($2-1$) by replacing the subscripts from 10 to 21
and using a new CO($2-1$)-to-H$_2$ conversion factor $\alpha_{21} \equiv \alpha_{10}/R_{\rm 21/10}$,
where $R_{\rm 21/10} [\equiv T_{21}/T_{10}]$ is the CO $J=2-1/1-0$ line ratio in brightness temperature.

In practice, $\alpha_{10}$ and $R_{\rm 21/10}$ are carried over in use of CO($J=2-1$)
as these are the parameters that have been measured.
Equation~(\ref{eq:mass_s10}) is now
\begin{equation}
M_{\rm mol} = \left( \frac{\alpha_{\rm 10}}{R_{\rm 21/10}} \right) \frac{c^2}{2 \nu_{21}^2 k} \left[ \int S_{21}dv \right] D^2. \label{eq:mass_s21}
\end{equation}

%\paragraph{Numerical Expressions}

A numerical evaluation gives
\begin{equation}
\left( \frac{M_{\rm H_2}}{M_{\odot}} \right) = 3.8 \times 10^{3} \left( \frac{\alpha_{\rm 10}}{4.3\,M_{\odot}\,{\rm pc^{-2} [K\cdot km/s]}^{-1}} \right) \left( \frac{R_{21/10}}{0.7} \right)^{-1} \left( \frac{ \int  S_{21}dv}{\rm Jy \cdot km/s} \right) \left( \frac{D}{\rm Mpc}\right)^2.\label{eq:m_s21_num}
\end{equation}

The typical GMC with $4\times10^5\,M_{\odot}$ and $dv=8.9\,\rm
km/s$ has $\int S_{21}dv\sim4.2\,\rm Jy \, km/s$ or $S_{21}\sim
470\,\rm mJy$ at $D=5\,\rm Mpc$.
Note $S_{21}>S_{10}$ for the same GMC
because $S_{21}/S_{10} = (\nu_{21}/ \nu_{10})^2 T_{21}/ T_{10}= (\nu_{21}/
\nu_{10})^2 R_{21/10}\sim2.8$ from Eq.~(\ref{eq:t_s}),
where the $(\nu_{21}/ \nu_{10})^2$ term arises from two facts:
at the higher frequency,
(a) each photon carries twice the energy, and 
(b) there are two times more photons in each frequency interval $d\nu$,
which is in the denominator of the definition of flux density $S$.
Empirically, $R_{\rm 21/10}\sim 0.7 $ on average in the MW (\citealt{Sakamoto:1997ys, Hasegawa:1997lr}),
which is consistent with a theoretical explanation under the conditions of the MW disk
(\citealt{Scoville:1974yu, Goldreich:1974jh}; see Sect.~\ref{sec:ISMextreme}).

%%%
\subsection{Observations of the Molecular ISM using Dust Continuum Emission}

Continuum emission from dust provides an alternative means for ISM mass measurement.
Dust is mixed in the gas phase ISM, and its emission at millimetre/submillimetre waves
correlates well with the fluxes of both atomic gas (H{\sc i} 21\,cm emission) and molecular gas (CO emission).
%Combined with a large bandwidth of continuum observations, the negative $k$-correction
%boosts the feasibility of dust continuum observations at high redshift galaxies
%in submillimetre/millimetre wavelength (REF - Scoville).
\citet{Scoville:2016aa} discussed the usage and calibration of dust emission for ISM mass
measurement.
We briefly summarize the basic equations, whose normalization will be adjusted with an empirical
fitting in the end.

The radiative transfer equation gives the brightness of dust emission
\begin{equation}
I_{\nu}=(1-e^{-\tau_{\nu}}) B_{\nu}(T_{\rm d})
\end{equation}
with the black body radiation $B_{\nu}(T_{\rm d})$ at the dust temperature $T_{\rm d}$ and the optical depth $\tau_{\nu}$.
The flux density of dust is an integration:
\begin{equation}
S_{\nu} = \int (1-e^{-\tau_{\nu}}) B_{\nu}(T_{\rm d}) d\Omega_{\rm B} = (1-e^{-\tau_{\nu}}) B_{\nu}(T_{\rm d}) \Omega_{\rm B},\label{eq:dustfluxdensity}
\end{equation}
where $B_{\nu}$ and $\tau_{\nu}$ are assumed constant within $\Omega_{\rm B}$ [$= \int d\Omega_{\rm B}$].
When the integration is over the beam area, $S_{\nu}$ is the flux density within the beam,
and $(S_{\nu}/\Omega_{\rm B})$, from Eq.~(\ref{eq:dustfluxdensity}), is in Jy/beam.

An integration of $S_{\nu}$ over the entire sky area at the distance of $D$ (i.e., $\int dA=D^2 \int_{4 \pi} d\Omega_{\rm A} = 4 \pi D^2$) gives the luminosity 
\begin{eqnarray}
L_{\nu} &=& \int (1-e^{-\tau_{\nu}}) B_{\nu}(T_{\rm d}) \Omega_{\rm B} dA = (1-e^{-\tau_{\nu}}) B_{\nu}(T_{\rm d}) \Omega_{\rm B} 4\pi D^2 \\
&\approx& 4 \pi \tau_{\nu} B_{\nu}(T_{\rm d}) D^2 \Omega_{\rm B} =
4 \pi \kappa_{\nu} \Sigma_{\rm d}  B_{\nu}(T_{\rm d}) D^2 \Omega_{\rm B} =4 \pi \kappa_{\nu} M_{\rm d} B_{\nu}(T_{\rm d}). \label{eq:dustlum}
\end{eqnarray}
The dust is optically thin at mm/sub-mm wavelengths, and we used
$(1-e^{-\tau_{\nu}}) \sim \tau_{\nu} = \kappa_{\nu} \Sigma_{\rm d}$,
where $\kappa_{\nu}$ and $\Sigma_{\rm d}$ are the absorption coefficient and surface density of dust.
The dust mass within the beam is $M_{\rm d} = \Sigma_{\rm d} D^2 \Omega_{\rm B}$.
Obviously, the dust continuum luminosity depends on the dust properties (e.g., compositions and
size distribution; via $\kappa_{\nu}$),
amount ($M_{\rm d}$), and temperature ($T_{\rm d}$).

Equation (\ref{eq:dustlum}) gives the mass-to-light ratio for
dust\index{dust mass-to-light ratio}

\begin{equation}
\frac{M_{\rm d}}{L_{\nu}} = \frac{1}{4 \pi \kappa_{\nu} B_{\nu}(T_{\rm d})}.\label{eq:dustml1}
\end{equation}
We convert $M_{\rm d}$ into gas mass, $M_{\rm gas}=\delta_{\rm GDR}M_{\rm d}$,
with the gas-to-dust ratio $\delta_{\rm GDR}$.
By re-defining the dust absorption coefficient $\kappa_{\nu}^{\prime} \equiv \kappa_{\nu}/\delta_{\rm GDR}$
(the absorption coefficient per unit total mass of gas), the gas mass-to-dust continuum flux
ratio $\gamma_{\nu}$ at the frequency $\nu$ becomes,
\begin{equation}
\gamma_{\nu} \equiv \frac{M_{\rm gas}}{L_{\nu}} = \frac{1}{4 \pi \kappa_{\nu}^{\prime} B_{\nu}(T_{\rm d})}. \label{eq:dustml}
\end{equation}
Once $\gamma_{\nu}$ is obtained, the gas mass is estimated as $M_{\rm gas}=\gamma_{\nu} L_{\nu}$.
Here, we use the character $\gamma$, instead of $\alpha$ that \citet{Scoville:2016aa} used,
to avoid a confusion with the CO-to-H$_2$ conversion factor.
Dust continuum emission is associated with H{\sc i} and H$_2$,
and $M_{\rm gas}\sim M_{\rm mol}$ in dense, molecule-dominated regions ($\gtrsim 100\,\rm cm^{-3}$).

%\paragraph{Numerical Evaluations}

The $\kappa_{\nu}^{\prime}$ can be approximated as a power-law
$\kappa_{\nu}^{\prime}=\kappa_{850\mu m}^{\prime} (\lambda/850\mu m)^{-\beta}$
with the spectral index $\beta \sim 1.8$ (\citealt{Planck-Collaboration:2011qy}) and
coefficient $\kappa_{850\mu m}^{\prime}$ at $\lambda=850\rm \mu m$ (352\,GHz).
In order to show the frequency dependence explicitly,
we separate $B_{\nu}(T_{\rm d})$ into the Rayleigh-Jeans term and the correction term $\Gamma_{\nu}(T_{\rm d})$
as $B_{\nu}(T_{\rm d})=(2\nu^2k T_{\rm d}/c^2) \Gamma_{\nu}(T_{\rm d})$, where 
\begin{equation}
\Gamma_{\nu}(T_{\rm d}) = \frac{ x}{e^x-1} \,\,\,\, \textrm{with} \,\,\,\,x=\frac{h\nu}{k T_{\rm d}}.
\end{equation}
Equation (\ref{eq:dustml}) has the dependence
$\gamma_{\nu} \propto  \nu^{-(\beta + 2)} T_{\rm d}^{-1} \Gamma_{\nu}(T_{\rm d})^{-1}$,
and the proportionality coefficient, including $\kappa_{850\mu m}^{\prime}$ and $\delta_{\rm GDR}$,
is evaluated empirically.

\citet{Scoville:2016aa} cautioned that $T_{\rm d}$ should not be derived from a spectral energy distribution fit (which gives
a luminosity-weighted average $T_{\rm d}$ biased toward hot dust
with a peak in the infrared). Instead, they suggested to use a mass-weighted $T_{\rm d}$
for the bulk dust component where the most mass resides.
\citet{Scoville:2016aa} adopted $T_{\rm d}=25\,\rm K$ and calibrated $\gamma_{\nu850\mu m}$
from an empirical comparison of $M_{\rm mol}$ (from CO measurements) and $L_{\nu}$,
\begin{equation}
\left( \frac{\gamma_{\nu}}{M_{\odot} {\rm [Jy\,cm^2]}^{-1}} \right) =1.5\pm 0.4 \times 10^{3} \left( \frac{\nu}{352\,\rm GHz} \right)^{-3.8} \left( \frac{T_{\rm d}}{25\,\rm K} \right)^{-1} \left( \frac{\Gamma_{\nu}(T_{\rm d})}{\Gamma_{\nu 850\mu  m}(25\,K)} \right)^{-1}.
\end{equation}
The luminosity is calculated from the observed $S_{\nu}$ in Jy and distance $D$ in centimetre
as $L_{\nu}=4 \pi D^2 S_{\nu}$ [$\rm Jy\,cm^2$]. The gas mass is then
$M_{\rm mol} = \gamma_{\nu} L_{\nu}$.

%%%
\subsection{The ISM in Extreme Environments Such as the
  Outskirts}\label{sec:ISMextreme}\index{extreme
  environment}\index{molecular mass: calibration issues}

The methods for molecular ISM mass measurement that we discussed above were developed
and calibrated mainly for the inner parts of galaxies.
However, it is not guaranteed that these calibrations are valid
in extreme environments such as galaxy outskirts.
In fact, metallicities appear to be lower in the outskirts than in the inner part (see Bresolin, this volume).
On a 1\,kpc scale average, gas and stellar surface densities,
and hence stellar radiation fields, are also lower, although it is not clear if these trends
persist at smaller scales, e.g., cloud scales, where the molecular ISM typically exists.
Empirically, $\alpha_{\rm 10}$ could be larger when metallicities are lower,
and $R_{\rm 21/10}$ could be smaller when gas density and/or temperature are lower.

In order to search for the molecular ISM and to understand star formation in the outskirts,
it is important to take into account the properties and conditions of the ISM there. Here we
explain some aspects that may bias measurements if the above equations
are applied naively as they are. These potential biases should not
discourage future research, and instead, should be adjusted
continuously as we learn more about the ISM in the extreme
environment.

\subsubsection{Variations of $\alpha_{\rm 10}$ (or $X_{\rm
    CO}$)}\index{$X_{\rm CO}$ variations}

The CO-to-H$_2$ conversion factor $\alpha_{\rm 10}$ (or $X_{\rm CO}$)
is a mass-to-light ratio between the CO($1-0$) luminosity and the
molecular ISM mass (\citealt{Bolatto:2013ys}).
Empirically, this factor increases with decreasing metallicity\index{metallicity} (\citealt{Arimoto:1996aa, Leroy:2011lr})
due to the decreasing abundance of CO over H$_2$.
At the low metallicity of the small Magellanic cloud ($\sim 1/10 \, Z_{\odot}$),
$\alpha_{\rm 10}$ appears $\sim10-20$ times larger (\citealt{Arimoto:1996aa, Leroy:2011lr}).

This trend can be understood based on the self-shielding nature of molecular clouds.
Molecules on cloud surfaces are constantly photo-dissociated by stellar UV radiation.
At high densities within clouds, the formation rate of molecules can be as fast as
the dissociation rate, and hence molecules are maintained in molecular clouds.
The depth where molecules are maintained depends on the strength of the
ambient UV radiation field and its attenuation by line absorptions by
the molecules themselves as well as by continuum absorption by dust
(\citealt{van-Dishoeck:1988br}).

\begin{figure}[h]
\sidecaption
\includegraphics[scale=.38]{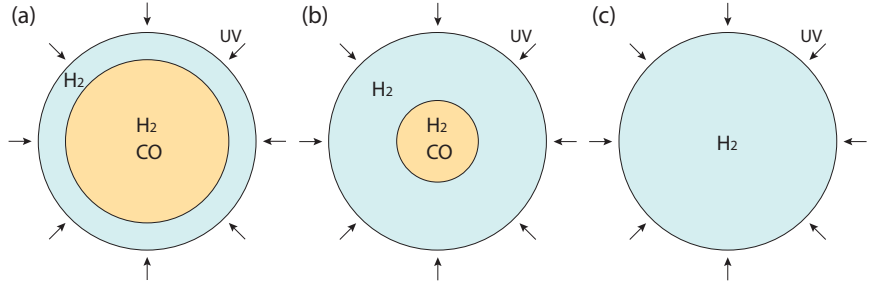}
\caption{Self-shielding nature of molecules in molecular clouds. The abundance of molecules is maintained
in clouds, since the destruction (photo-dissociation by UV radiation) and formation rates are in balance.
The shielding from ambient UV radiation is mainly due to line absorption by molecules themselves.
Therefore, the abundant H$_2$ molecules become optically thick at the absorption line wavelengths
on the skin of clouds, while UV photons for CO dissociation can get deeper into clouds.
This mechanism generates the CO-dark H$_2$ layer on the surface of molecular clouds.
This layer can become thicker ({\it panels a, b, c}) under several conditions: e.g.,
lower metallicity or stronger local radiation field.
The CO-to-H$_2$ conversion factor $\alpha_{\rm 10}$ (or $X_{\rm CO}$) increases
with the increasing thickness of the CO-dark H$_2$ layer, and
therefore, with lower metallicity or stronger local radiation field}
\label{fig:codark} 
\end{figure}

H$_2$ is $\sim10^4$ times more abundant than CO. It can easily become optically thick on
the skin of cloud surfaces and be self-shielded (Fig.~\ref{fig:codark}).
On the other hand, UV photons for CO dissociation penetrate deeper into the cloud
due to its lower abundance.
This process generates the CO-dark H$_2$\index{CO-dark H$_2$} layer around molecular clouds (Fig.~\ref{fig:codark}b; \citealt{Wolfire:2010fk}).
Shielding by dust is more important for CO than H$_{2}$. Therefore, if
the metallicity or dust abundance is low, the UV photons for CO
dissociation reach deeper and deeper, and eventually destroy all CO
molecules while H$_2$ still remains (Fig.~\ref{fig:codark}c).
As the CO-dark H$_2$ layer becomes
thicker, $L_{10}$ decreases while $M_{\rm H_2}$ stays high, resulting
in a larger $\alpha_{\rm 10}$ in a low metallicity environment, such
as galaxy outskirts. Since this process depends on the depth that
photons can penetrate (through dust attenuation as well as line absorption),
the visual extinction $A_{\rm V}$ is often used
as a parameter to characterize $\alpha_{\rm 10}$ (or $X_{\rm CO}$).

\subsubsection{Variations of $R_{\rm
    21/10}$}\index{CO($J=2-1$)/CO($J=1-0$) variations}

The CO($2-1$) line emission is useful to locate
the molecular ISM and to derive a rough estimation of its
mass. However, the higher transitions inevitably suffer from
excitation conditions\index{excitation conditions}. Indeed, $R_{\rm
21/10}$ ($\equiv T_{\rm 21}/T_{\rm 10}$) has been observed to vary by
a factor of $2-3$ in the MW and in other nearby galaxies, e.g., between
star-forming molecular clouds (typically $R_{\rm 21/10}\sim 0.7-1.0$
and occasionally up to 1.2) and dormant clouds ($\sim0.4-0.7$), and
between spiral arms ($>0.7$) and inter-arm regions ($<0.7$;
\citealt{Sakamoto:1997ys, Koda:2012lr}).  The variation may be negligible
for finding molecular gas, but may cause a systematic bias, for
example, in comparing galaxy outskirts with inner disks. It is
noteworthy that $R_{\rm 21/10}$ changes systematically with star
formation activity, and varies along the direction of the
Kennicutt-Schmidt relation, which can introduce a bias.

\begin{figure}[t]
\sidecaption
\includegraphics[scale=.6]{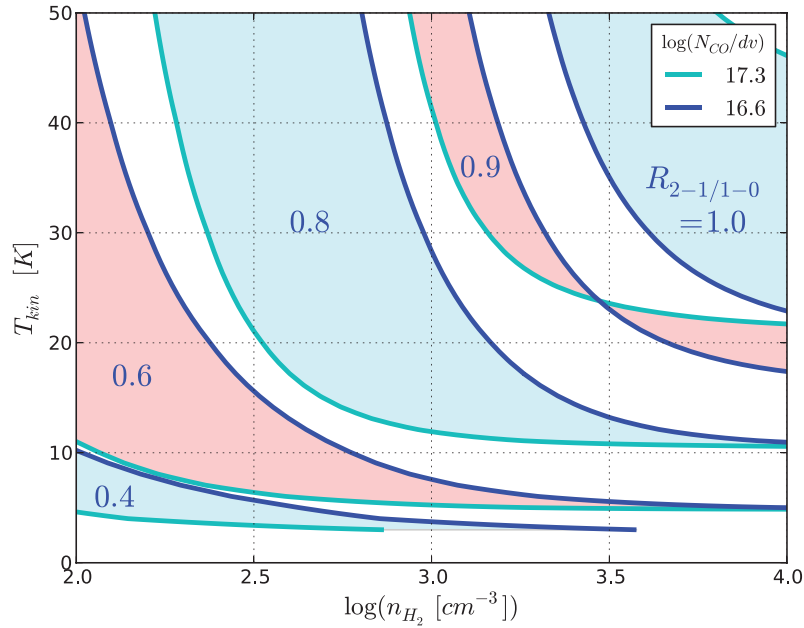}
\caption{The CO $J=2-1/1-0$ line ratios as function of the gas kinetic temperature $T_{\rm kin}$
and H$_2$ density $n_{\rm H2}$ under the LVG approximation (from \citealt{Koda:2012lr}).
Most GMCs in the MW have CO column density in the range of $\log(N_{\rm CO}/dv)\sim16.6$ to 17.3,
assuming the CO fractional abundance to H$_2$ of $8\times 10^{-5}$.
An average GMC in the MW has $n_{\rm H_2}\sim 300 \,\rm cm^{-3}$ and $T_{\rm k}\sim10 \,\rm K$,
and therefore shows $R_{\rm 21/10}\sim$0.6-0.7. $R_{\rm 21/10}$ is $<0.7$ if the density and/or temperature
decrease by a factor of $2-3$, and $R_{\rm 21/10}$ is $>0.7$ if the
density and/or temperature increase by a factor of $2-3$.
Observationally, dormant clouds typically have $R_{\rm 21/10}=0.4-0.7$,
while actively star forming clouds have $R_{\rm 21/10}=0.7-1.0$ (and occasionally up to $\sim1.2$; \citealt{Sakamoto:1997ys, Hasegawa:1997lr}).
There is also a systematic variation between spiral arms ($R_{\rm
  21/10}>0.7$) and interarm regions ($R_{\rm 21/10}<0.7$; \citealt{Koda:2012lr})}
\label{fig:co2110} 
\end{figure}

Theoretically, $R_{\rm 21/10}$ is controlled by three parameters: the volume density $n_{\rm H_2}$
and kinetic temperature $T_{\rm k}$ -- which determine the CO excitation condition due to
collisions -- and the column density $N_{\rm CO}$, which controls radiative transfer and
photon trapping (\citealt{Scoville:1974yu, Goldreich:1974jh}).
Figure~\ref{fig:co2110} shows the variation of $R_{\rm 21/10}$ with respect to $n_{\rm H_2}$
and $T_{\rm k}$ under the large velocity gradient (LVG) approximation.
In this approximation, the Doppler shift due to a cloud's internal velocity gradient
is assumed to be large enough such that any two parcels along the line
of sight do not overlap in velocity space. The front
parcel does not block emission from the back parcel, and the optical
depth is determined only locally within the parcel (or in small
$dv$). Therefore, the column density is expressed per velocity $N_{\rm CO}/dv$.
A typical velocity range in molecular clouds is adopted for this figure.
An average GMC in the MW has $n_{\rm H_2}\sim 300 \,\rm cm^{-3}$ and $T_{\rm k}\sim10 \,\rm K$
(\citealt{Scoville:1987vo}), which results in $R_{\rm 21/10}$ of $\sim0.6-0.7$.
If the density and/or temperature is a factor of $2-3$ higher due to a contraction before star formation
or feedback from young stars, the ratio increases to $R_{\rm 21/10}>0.7$. On the contrary,
if a cloud is dormant compared to the average, the ratio is lower $R_{\rm 21/10}<0.7$.

In the MW, cloud properties appear to change with the galactocentric radius (\citealt{heyer15}).
If their densities or temperatures are lower in the outskirts, it would result in a lower $R_{\rm 21/10}$,
and hence, a higher H$_2$ mass at a given CO($2-1$) luminosity.
If the $R_{\rm 21/10}$ variation is not accounted for, it could result
in a bias when clouds within the inner disk and in the outskirts are compared.

\subsubsection{Variations of Dust Properties and Temperature}\index{dust
  emission variations}

The gas mass-to-dust luminosity $M_{\rm gas}/L_{\nu}$ depends on the
dust properties/emissivity ($\kappa_{\nu}$), dust temperature ($T_{\rm
d}$), and gas-to-dust ratio ($\delta_{\rm GDR}$) -- see Eqs.~(\ref{eq:dustml1}) and (\ref{eq:dustml}). All of these parameters
could change in galaxy outskirts, which have low average
metallicity\index{metallicity}, density, and stellar radiation
field. Of course, the assumption of a single $T_{\rm d}$ casts a
limitation to the measurement as the ISM is 
multi-phase in reality, although the key idea of using
Eqs.~(\ref{eq:dustml1}) and (\ref{eq:dustml}) is to target regions
where the cold, molecular ISM is dominant (\citealt{Scoville:2016aa}).
The $\delta_{\rm GDR}$ may increase with decreasing metallicity by about an order of magnitude
($\delta_{\rm GDR} \sim 40 \rightarrow$ 400) for the change of metallicity $12+\log({\rm O/H})$ 
from $\sim 9.0 \rightarrow$ 8.0 (their Fig.~6; \citealt{Leroy:2011lr}).
If this trend applies to the outskirts, Eq.~(\ref{eq:dustml}) would tend to underestimate the gas mass
by up to an order of magnitude.

Excess dust emission at millimetre/submillimetre wavelengths has been
reported in the small and large Magellanic clouds (SMC and LMC) and
other dwarfs (\citealt{Bot:2010aa, Dale:2012aa}; although see also
\citealt{Kirkpatrick:2013aa}). This excess emission appears
significant when spectral energy distribution fits to infrared data
are extrapolated to millimetre/submillimetre wavelengths. Among the
possible explanations are the presence of very cold dust, a change of
the dust spectral index, and spinning dust emission (e.g., \citealt{Bot:2010aa}).
\citet{Gordon:2014aa} suggested that variations in the dust emissivity
are the most probable cause in the LMC and SMC
from their analysis of infrared data from the {\it Herschel Space Observatory}.
The environment of galaxy outskirts may be similar to those of the LMC/SMC.
The excess emission (27\% and 43\% for the LMC and SMC, respectively;
\citealt{Gordon:2014aa}) can be ignored if one only needs to locate
dust in the vast outskirts, but could cause a systematic bias when the
ISM is compared between inner disks and outskirts. 

\section{Molecular Gas Observations in the Outskirts of Disk Galaxies}\label{sec:disks}

A primary motivation for molecular gas observations in the outskirts of
disk galaxies has been to study molecular clouds and star formation in
an extreme environment with lower average density and
metallicity. Many researchers highlight that these studies may teach
us about the early Universe, where these conditions were more
prevalent.

\subsection{The Milky Way}\index{Milky Way: molecular gas}
The MW is the disk galaxy with the most molecular gas detections in
the outskirts, with pioneering studies of the outer disk
molecular gas and star formation properties beginning in the 1980s
(e.g., \citealt{fich84, brand88}). The MW
can serve as a model for the types of studies that can be done in
nearby galaxies with larger and more sensitive facilities.  We will
use ``outer'' MW to refer to galactocentric radii between the solar
circle ($R_{\rm Gal} > R_{\odot} = 8.5 \, {\rm kpc}$) and the edge of
the optical disk, which is estimated to be at $R_{\rm Gal} \sim 13-19
\, {\rm kpc}$ (\citealt{ruffle07,sale10} and references therein). We
will use ``outskirts'' to refer to galactocentric radii beyond the
edge of the optical disk.

\begin{figure}
\begin{center}
\includegraphics[width=\textwidth]{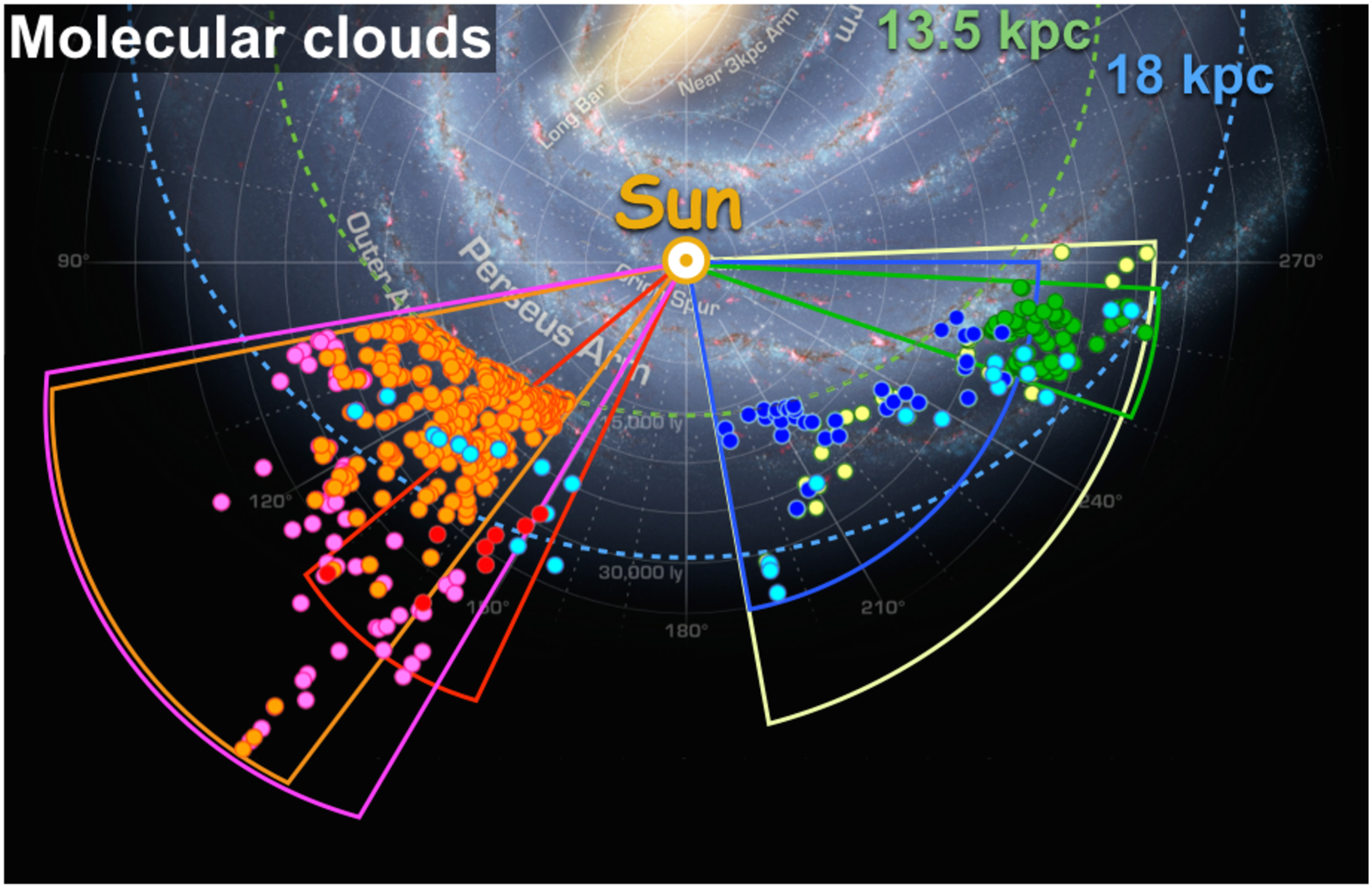}
\caption{Figure from N. Izumi (personal communication) showing the known
molecular clouds at $R_{\rm Gal} > 13.5 \, {\rm kpc}$ in the second and third
quadrants overlaid on an artist's conception of the MW (R. Hurt:
NASA/JPL-Caltech/SSC). The colours correspond to the following surveys: 
orange: \citet{brunt03}, magenta: \citet{sun15}, red: \citet{digel94},
cyan: \citet{brand94}, blue: \citet{may97}, green:
\citet{nakagawa05}, yellow: \citet{vazquez08}. The points represent
molecular clouds and the fan-shaped regions represent the survey
area. The distances were derived assuming $R_{\odot} = 8.5 \, {\rm
kpc}$ and a solar orbital speed of $V_{\odot} = 220 \, {\rm km \, s^{-1}}$}
\label{fig:izumi}
\end{center}
\end{figure}

Only about $2\%$ of the molecular mass of the MW is at $R_{\rm Gal}
> 14.5 \, {\rm kpc}$ (\citealt{nakagawa05} estimated the molecular mass
at $R_{\rm Gal}>14.5 \, {\rm kpc}$ to be $2 \times 10^7 \, M_{\odot}$ while
\citealt{heyer15} estimated the total molecular mass of the Galaxy to
be $(1 \pm 0.3) \times 10^{9} \, M_{\odot}$). N. Izumi (personal communication)
collected the known molecular clouds with $R_{\rm Gal}>13.5 \, {\rm
kpc}$ in the second and third quadrants (Fig.~\ref{fig:izumi}). The
molecular cloud with the largest known galactocentric radius is
probably Digel Cloud 1 with a kinematic galactocentric radius of
$R_{\rm Gal} = 22 \, {\rm kpc}$, dynamical mass of $\sim 6 \times
10^{4} \, M_{\odot}$, and radius of $36 \, {\rm pc}$ (Digel
Cloud 2 has a larger kinematic distance of $R_{\rm Gal} = 24 \, {\rm
kpc}$, but the photometric distance is $R_{\rm Gal} = 15-19 \, {\rm
kpc}$ based on optical spectroscopy of an associated B
star; \citealt{digel94,yasui06,yasui08,izumi14}). Digel Cloud 1 is beyond 
the edge of the optical disk but well within the H{\sc i} disk, which extends to 
$R_{\rm Gal} \sim 30 \, {\rm kpc}$ (\citealt{digel94,ruffle07} and references
therein).

Extremely tenuous H$_2$ gas is mixed with the H{\sc i} gas in the Galactic
halo with a fraction of H$_2$ over H{\sc i} of only $10^{-4 \sim -5}$
(\citealt{lehner02}). Such tenuous H$_2$ is observed via UV 
absorption, e.g., toward the Magellanic stream (\citealt{lehner02}) and high
velocity clouds (HVCs; \citealt{bluhm01}). This component is important
for understanding the complex physics of the ISM, but is not a major
molecular component in galaxy outskirts. We therefore do not discuss
this component further in this review.

\subsubsection{Properties of Molecular Clouds in the Outer Milky
  Way}\index{molecular cloud properties}

In this Section we highlight studies that have compared the mass, size,
and mass surface density of molecular clouds in the outer MW to clouds
in the inner MW. Molecular clouds are the site of star formation, and hence,
comparisons of their properties between the inner and outer MW is important.
In general, molecular clouds in the outer MW have lower mass and mass
surface density than clouds in the inner disk. We also describe how
molecular clouds have been used to trace spiral arms into the
outskirts and to study relatively high-mass star formation.

\citet{heyer15} combined published data on the CO surface brightness
out to $R_{\rm Gal} \sim 20 \, {\rm kpc}$. The clouds in the outer
MW and outskirts are $\sim 7$ times fainter than clouds in the
inner MW (and even fainter relative to the Galactic
centre). Assuming a constant $X_{\rm CO}$, this corresponds to a
factor of $\sim 7$ decrease in the mass surface density of molecular
clouds. \citet{heyer15} argued that there is a real decrease in the
mass surface density of the molecular clouds, perhaps caused by the
lower mid-plane pressure or stronger local FUV radiation field in the
outer Galaxy. However, there is also evidence that the outer MW
requires a larger $X_{\rm CO}$ to convert the CO surface 
brightness into the mass surface density (see Sect.\ref{sec:ISMextreme}). Therefore the mass surface
density likely decreases by somewhat less than a factor of $\sim 7$.

The mass function of molecular clouds in the outer MW ($9.5 \,
{\rm kpc} \lesssim R_{\rm Gal} \lesssim 13.5 \, {\rm kpc}$ in this
study) has a steeper power law index than that in the inner MW, such
that the outer disk hosts more of its molecular mass in lower-mass
clouds (\citealt{rosolowsky05}, based on the $330 \, {\rm deg^2}$
\citealt{heyer98} catalogue and analysis in \citealt{heyer01} and
\citealt{brunt03}), although this conclusion may at some level be a
result of variable angular resolution (\citealt{heyer15}). The mass
function of the outer MW shows no clear evidence for a
truncation at the high-mass end, but under some assumptions
\citet{rosolowsky05} estimated that the maximum molecular cloud mass
is $\sim2-3 \times 10^{5} \, M_{\odot}$. In contrast,
\citet{rosolowsky05} concluded that the inner MW shows a clear
truncation with maximum molecular cloud mass of $\sim 3 \times 10^{6}
\, M_{\odot}$. Because of the small number of known clouds, the
apparent lack of massive clouds in the outer MW might be due to a
sampling effect. This possibility should be addressed in future
studies, as a truncation, if it exists, would be an important clue to
understanding cloud physics in the outskirts.

\citet{heyer01} concluded that the size distribution of molecular
clouds in the outer MW is similar to the distribution in the
inner MW from \citet{solomon87}, but note that surveys
with fewer clouds and different galactocentric distance ranges reached
different conclusions. \citet{may97} concluded that outer MW
clouds have smaller sizes than the inner MW while
\citet{brand95} concluded that the outer MW clouds have larger
sizes than inner MW clouds at the same mass. While there are
conflicting results in the literature, it seems natural to conclude
that an outer MW cloud must have a larger radius than an inner MW
cloud at the same mass because it appears that the mass surface
density of clouds is lower in the outer MW (see above and
\citealt{heyer15}).

Molecular gas observations in the outskirts of the MW have been
used to identify spiral arms\index{spiral arms}. \citet{dame11}
discovered a spiral arm in the first quadrant at $R_{\rm Gal} \sim 15
\, {\rm kpc}$, based on H{\sc i} and CO data. Their new arm is consistent
with being an extension of the Scutum-Centaurus arm. \citet{sun15}
also used H{\sc i} and CO data to discover an arm in the second quadrant at
$R_{\rm Gal} = 15 - 19 \, {\rm kpc}$. This arm could be a further
continuation of the Scutum-Centaurus arm and the \citet{dame11}
arm. These kinds of studies are important not only to map the spiral
structure of the MW, but also to help understand the observation that
star formation in the outskirts of other galaxies often follows spiral 
arms.

Another important goal of molecular gas studies in the outskirts of
the MW has been to understand the connection with star
formation\index{star formation: Milky Way} under low density and
metallicity conditions. For example, \citet{brand07} studied an
IRAS-selected molecular cloud with a mass of $4.5 - 6.6 \times 10^{3}
\, M_{\odot}$ at $R_{\rm Gal} \sim 20.2 \, {\rm kpc}$. They discovered
an embedded cluster of 60 stars and the lack of radio continuum
emission limits the most massive star to be later than B0.5. In
addition, \citet{kobayashi08} studied Digel Cloud 2, which is really
two clouds each with a mass of $\sim 5 \times 10^{3} \,
M_{\odot}$. They discovered embedded clusters in each of the 
clouds. One cluster likely contains a Herbig Ae/Be star and there are
also several Herbig Ae/Be star candidates, a B0-B1 star, and an H{\sc ii}
region nearby. Therefore, high-mass star formation has occurred near
this low-mass molecular cloud. We encourage more study on the
relationship between cloud mass and the most massive star
present, as extragalactic studies can trace O and B stars relatively
easily, but have difficulty detecting the parent molecular clouds (see 
Sect.~\ref{sec:disk_galaxies_mol_gas}).

In the outskirts of the MW and other galaxies, it is important
to ask what triggers molecular cloud and star formation. In Digel
Cloud 2, star formation may have been triggered by the expanding H{\sc i}
shell of a nearby supernova remnant
(\citealt{kobayashi00,yasui06,kobayashi08}) while \citet{izumi14}
hypothesized that the star formation in Digel Cloud 1 may have been
triggered by interaction with a nearby HVC.

\subsection{Extragalactic Disk Galaxies}\index{disk galaxies}

We can study molecular gas in more varied environments by moving from
the MW to extragalactic disk galaxies. In this Section, we use
``outskirts'' to refer to galactocentric radii greater than the
optical radius ($R_{\rm Gal}>r_{25}$).

\subsubsection{Molecular Gas Detections}\index{molecular clouds: extragalactic}
\label{sec:disk_galaxies_mol_gas}

Numerous attempts to detect CO beyond the optical radius in the disks
of spiral galaxies have failed, although many of the non-detections
are unpublished (\citealt{watson16,morokuma16}; J.\ Braine, F.\ Combes,
J.\ Donovan Meyer, and A.\ Gil de Paz, personal communications). To our
knowledge, there are only four isolated spiral galaxies
with published CO detections beyond the optical radius
(\citealt{braine04,braine07,braine10,braine12,dessauges14}). Table~\ref{tab:disk_galaxies} 
summarizes the number of detected regions and their range of
galactocentric radii and molecular gas masses. Extragalactic studies
have not yet reached the molecular gas masses that are typical in the
outskirts of the MW ($2-20 \times 10^{3} \, M_{\odot}$
for the eleven Digel clouds at $R_{\rm Gal} = 18-22 \, {\rm kpc}$;
\citealt{digel94,kobayashi08}; see also \citealt{braine07}).

\begin{table}
\caption{Extragalactic disk galaxies in relative isolation with
  CO detections beyond the optical radius
  (\citealt{braine04,braine07,braine10,braine12,dessauges14}). 
  For M33, the molecular gas mass is for one of the detected clouds. For M63,
  the molecular gas mass is based on a sum of the CO line intensities in
  twelve pointings, two of which are detections. The NGC~4414,
  NGC~6946, and M63 masses were computed assuming $X_{\rm
  CO} = 2 \times 10^{20} \, {\rm cm^{-2} (K \, km \, s^{-1})^{-1}}$.}
\label{tab:disk_galaxies}       % Give a unique label
\begin{tabular}{p{1.5cm}p{1.5cm}p{2cm}p{1.5cm}p{4.5cm}}
\hline\noalign{\smallskip}
Galaxy & Detected & Galactocentric & Molecular & Method used for Mass \\
 & Regions   & Radius & Gas Mass &  \\
 & (\#)  & ($r_{25}$) & ($10^{5} \, M_{\odot}$) & \\
\noalign{\smallskip}\svhline\noalign{\smallskip}
NGC~4414 & 4 & $1.1 - 1.5$ & $10-20$   & Within 21'' IRAM $30 \, {\rm m}$ beam \\
NGC~6946 & 4 & $1.0 - 1.4$ & $1.7-3.3$ & Within 21'' IRAM $30 \, {\rm m}$ beam \\
M33           & 6 & $1.0 - 1.1$ & $0.43$      & Virial mass using resolved PdBI data \\
M63           & 2  & $1.36$       & $7.1$       & Sum of 12 IRAM $30 \, {\rm m}$ pointings \\
\noalign{\smallskip}\hline\noalign{\smallskip}
\end{tabular}
%$^a$ Table foot note (with superscript)
\end{table}

It would be useful to be able to predict where CO will be detected in
the outskirts of disk galaxies, both as a test of our understanding of
the physics of CO formation and destruction in extreme conditions (see
Sect.~3.4) and to help us efficiently collect more detections. Most of
the published CO studies selected high H{\sc i} column density regions or
regions near young stars traced by H$\alpha$, FUV, or FIR
emission. None of these selection methods is completely
reliable. \citet{braine10} concluded that CO is often associated with
large H{\sc i} and FIR structures, but it is not necessarily located at H{\sc i},
FIR, or H$\alpha$ peaks. Many factors might affect the association
between H{\sc i}, CO and star formation tracers. For example, the star forming
regions may drift away from their birthplaces over the $10-100 \, {\rm
  Myr}$ timescales traced by H$\alpha$, FUV, and FIR emission. In
addition, feedback from massive stars might destroy molecular clouds
more easily in the low-density outskirt environment. Finally, higher-resolution H{\sc i} maps may show better correlation with CO
emission. Sensitive, large-scale ($> {\rm kpc}$) maps of the outskirts
of disk galaxies may allow for a more impartial study of the
conditions that maximize the CO detection rate.

\subsubsection{Star Formation in Extragalactic Disk
  Galaxies}\index{star formation: extragalactic disk galaxies}

It is generally accepted that stars form from molecular gas
(e.g., \citealt{fukui10}) and that an important stage before star formation is
the conversion of H{\sc i} to H$_{2}$ (e.g., \citealt{leroy08}). A main tool to
study the connection between gas and star formation is the
Kennicutt-Schmidt law\index{Kennicutt-Schmidt law}
(\citealt{schmidt59,kennicutt98}), which is an empirical relationship
between the star formation rate (SFR) surface density ($\Sigma_{\rm
SFR}$) and the gas surface density. Within the optical 
disk of spiral galaxies, there is an approximately linear correlation
between $\Sigma_{\rm SFR}$ and the molecular hydrogen surface density
($\Sigma_{\rm H_{2}}$) but no correlation between $\Sigma_{\rm SFR}$
and the atomic hydrogen surface density ($\Sigma_{\rm HI}$; e.g.,
\citealt{bigiel08,schruba11}).

The majority of the published work connecting the SFR and gas density
in the outskirts of disk galaxies has focused on the atomic gas
because molecular gas is difficult to detect
(Sect.~\ref{sec:disk_galaxies_mol_gas}) and because the ISM is
dominantly atomic in the outskirts, at least on $\gtrsim \, {\rm kpc}$
scales. \citet{bigiel10} concluded that there is a correlation between
the FUV-based $\Sigma_{\rm SFR}$ and $\Sigma_{\rm HI}$ in the
outskirts of 17 disk galaxies and 5 dwarf galaxies. They measured a
longer depletion time in the outskirts, such that it will take on average
$10^{11}$ years to deplete the H{\sc i} gas reservoir in the outskirts
versus $10^{9}$ years to deplete the ${\rm H_{2}}$ gas reservoir
within the optical disk. \citet{roychowdhury15} reached a similar
conclusion using H{\sc i}-dominated regions in disks and dwarfs, including
some regions in the outskirts, although they concluded that the depletion
time is somewhat shorter than in the outskirts of the \citet{bigiel10} 
sample (see also \citealt{boissier07,dong08,barnes12}). The correlation
between $\Sigma_{\rm SFR}$ and $\Sigma_{\rm HI}$ is surprising because
there is no correlation within the optical disk. \citet{bigiel10}
suggested that high H{\sc i} column density is important for determining
where stars will form in the outskirts.

The study of the connection between molecular gas and star formation
in the outskirts has been limited by the few molecular gas
detections. Figures~5 and 6 show the relationship between $\Sigma_{\rm
SFR}$ and $\Sigma_{\rm H_{2}}$ for the molecular gas detections from
Table~\ref{tab:disk_galaxies} plus a number of deep CO upper
limits. In both panels the SFR was computed based on FUV and $24 \,
{\rm \mu m}$ data to account for the star formation that is unobscured
and obscured by dust.

\begin{figure}[b]
\begin{center}
\includegraphics[width=0.8\textwidth]{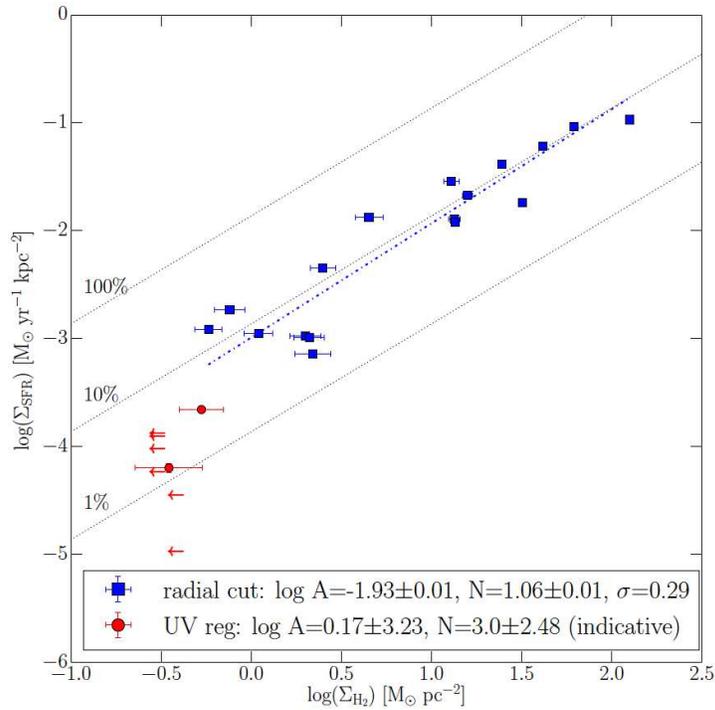}
\end{center}
\caption{Figure 7 from \citet{dessauges14} showing
the molecular-hydrogen Kennicutt-Schmidt relation for the star forming
regions in the UV-complex at $r=1.36 \, r_{25}$ in M63 (red points)
compared to regions within the optical disk (blue points). The blue
line shows the fit for the optical disk. The black lines represent
constant star formation efficiency, assuming a timescale of $10^{8}$
years. Credit: \citet{dessauges14}, reproduced with permission \copyright\ ESO}
\end{figure}

\begin{figure}[b]
\begin{center}
\includegraphics[width=0.8\textwidth]{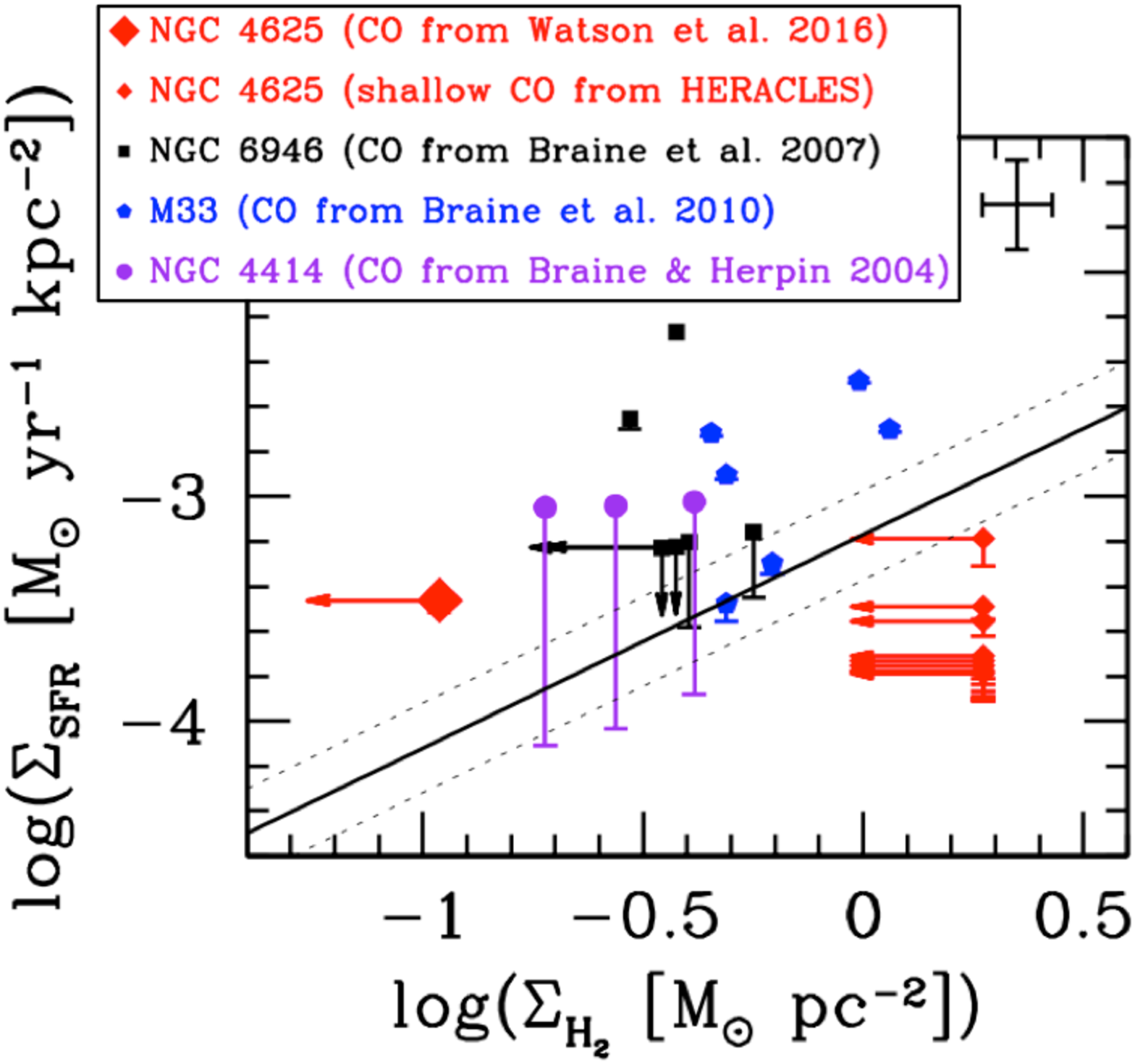}
\end{center}
\caption{The molecular hydrogen Kennicutt-Schmidt relation
for the remaining star 
forming regions that are beyond the optical radius in isolated
extragalactic disk galaxies and have published CO detections or
deep upper limits. The solid line shows the fit for the optical disk
of normal spiral galaxies at $\sim$kpc resolution, with the $1\sigma$
scatter shown by the dotted lines (\citealt{leroy13}). This figure was
originally presented in Fig.~4 in \citet{watson16}}
\end{figure}

\citet{dessauges14} studied a UV-bright region at $r=1.36 \, r_{25}$
in the XUV disk of M63 (Fig.~5). They detected CO in two out of
twelve pointings and concluded that the molecular gas has a low star
formation efficiency\index{star formation efficiency} (or,
equivalently, the molecular gas has a long depletion time) compared to
regions within the optical disk. They suggested that the low star
formation efficiency may be caused by a warp or by high
turbulence. \citet{watson16} measured a deep CO upper 
limit in a region at $r=3.4 \, r_{25}$ in the XUV disk of NGC~4625 and
compiled published CO measurements and upper limits for 15 regions in
the XUV disk or outskirts of NGC~4414, NGC~6946, and M33 from 
\citet{braine04} and \citet{braine07,braine10} (see Table~\ref{tab:disk_galaxies} 
and Fig.~6). They concluded that star-forming regions in the
outskirts are in general consistent with the same $\Sigma_{\rm
  SFR}$-$\Sigma_{\rm H_{2}}$ relationship that exists in the optical
disk. However, some points are offset to high star formation
efficiency (short depletion time), which may be because the authors selected
H$\alpha$- or FUV-bright regions that could have already exhausted
some of the molecular gas supply (as in \citealt{schruba10,kruijssen14}).

We should ask what stimulates the formation of molecular gas and stars
in the outskirts of disk galaxies. \citet{thilker07} suggested that
interactions may trigger the extended star formation in XUV disks
while \citet{holwerda12} suggested that cold accretion may be more
important. \citet{bush08,bush10} carried out hydrodynamic simulations
and concluded that spiral density waves can raise the density in an
extended gas disk to induce star formation (see also Sect.~4.1.1. of Debattista et al., this volume).

The state-of-the-art data from SINGS (\citealt{kennicutt03}), the
{\it GALEX} Nearby Galaxy Survey (\citealt{gildepaz07}), THINGS
(\citealt{walter08}), and HERACLES (\citealt{leroy09}) brought new insight
into the Kennicutt-Schmidt law within the optical disk of
spirals. Deeper CO surveys over wider areas in the outskirts could
bring a similar increase in our understanding of star formation at the
onset of the H{\sc i}-to-H$_{2}$ transition. In such wide-area studies, one
should keep in mind that the ``standard" physical condition of gas in
inner disks could change in the outskirts, which could affect the
measurements (Sect.~\ref{sec:ISMextreme}).

\subsubsection{Theory}\index{theory}

This Chapter focuses on observations, but here we briefly highlight
theoretical works that are related to molecular gas in the
outskirts. The majority of the relevant theoretical studies have
concentrated on the origin of gas in the outskirts
(e.g., \citealt{dekel06,sancisi08,sanchez14,mitra15}) and star
formation in the outskirts
(\citealt{bush08,bush10,ostriker10,krumholz13,sanchez14}; 
see also \citealt{roskar10,khoperskov15}). \citet{krumholz13} is
particularly relevant because he extended earlier work to develop an
analytic model for the atomic and molecular ISM and star formation in
outer disks. Krumholz assumed that hydrostatic equilibrium sets the
density of cold neutral gas in the outskirts and was able to match
the \citet{bigiel10} observations that show a correlation between
$\Sigma_{\rm SFR}$ and $\Sigma_{\rm HI}$ (see also Sect.~7 of Elmegreen and Hunter, this volume).

\section{Molecular Gas Observations in the Outskirts of Early-Type
  Galaxies}\index{early-type galaxies}\label{sec:ellipticals}
\label{sec:early_types}

Early-type galaxies were historically viewed as ``red and dead,'' with
little gas to form new stars. However, more recent surveys have found
reservoirs of cold gas both at galaxy centres and in the
outskirts. Molecular gas in the centres of early-type galaxies can
have an internal and/or external origin while the molecular gas in the 
outskirts often originated in a gas-rich companion that has interacted
or merged with the early-type. As in all of the environments we have
explored, stimuli can also trigger new molecule formation in the
outskirts of early-types.

We start with a review of H{\sc i} in the inner and outer regions of
early-type galaxies to put the molecular gas observations in
context. The ATLAS$^{\rm 3D}$ survey detected H{\sc i} in 32\% of 166
early-type galaxies in a volume-limited sample, down to a $3\sigma$
upper limit of $M_{\rm HI} = 5 \times 10^{6} - 5 \times 10^{7} \,
{M_{\odot}}$. Atomic gas in the outskirts of early-type galaxies is
even relatively common, as 14\% of the ATLAS$^{\rm 3D}$ sample have H{\sc i}
that extends out to more than 3.5 times the optical effective radius 
(\citealt{serra12}).

Most surveys of molecular gas in early-type galaxies have focussed on
the inner regions. 22\% of 260 early-type galaxies in the ATLAS$^{\rm
3D}$ sample were detected in CO, down to a $3\sigma$ upper limit of
$M_{\rm H_{2}} \sim 10^{7} - 10^{8} \, {M_{\odot}}$ (\citealt{young11};
see also \citealt{sage89,knapp96,welch03,combes07,welch10}). Within
the areas searched, the molecular gas is generally confined to the
central few kpc and is distributed in disks, bars plus rings, spiral
arms, or with a disrupted morphology (\citealt{young02, welch03, young08,
davis13, alatalo13}).

One important motivation for studies of molecular gas in early-type
galaxies has been to determine whether the gas is of internal or
external origin. Some of the molecular gas has likely either
been present since the galaxies transitioned to being early-type or
has accumulated from stellar mass loss (\citealt{faber76, young02,
  young08, mathews03, ciotti10}). In contrast, some molecular gas has
likely been accreted more recently through minor mergers and/or cold
accretion. This external origin is most clearly exhibited by galaxies
that display a misalignment between the kinematic axes of the
molecular/ionized gas and the stars (\citealt{young08, crocker08,
davis11, alatalo13}). In particular, \citet{alatalo13} concluded that 15
galaxies out of a sample of 40 show a kinematic misalignment of at
least 30 degrees, which is consistent with gas accretion via minor
mergers.

\begin{figure}
\begin{center}
\includegraphics[width=0.9\textwidth, angle=-90]{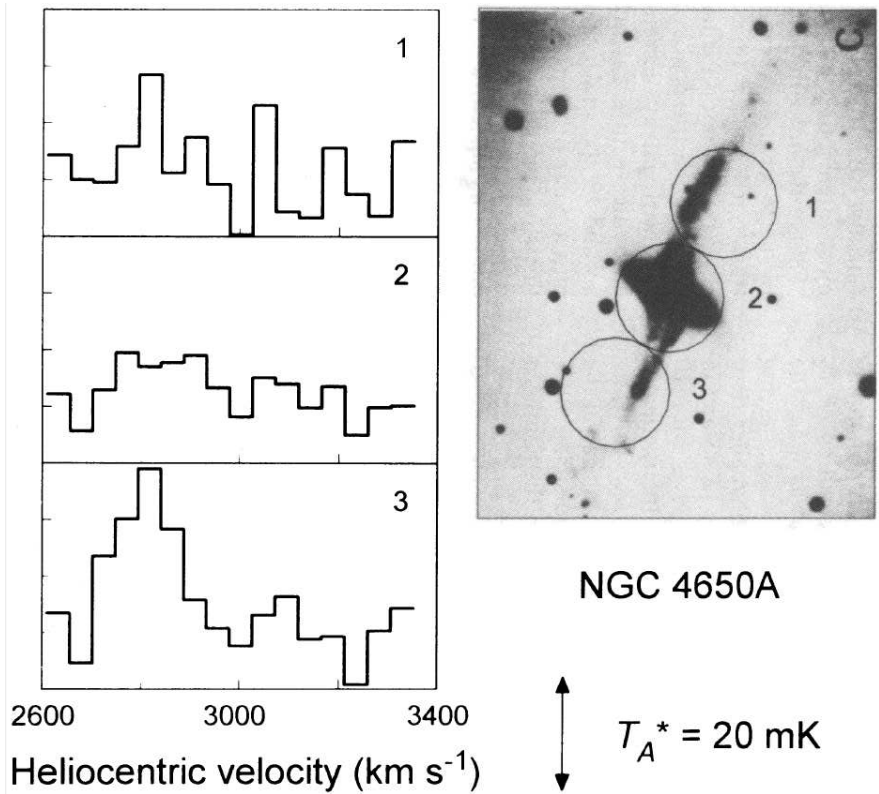}
\caption{Figure~2 from \citet{watson94} showing the Caltech
Submillimeter Observatory CO($2-1$) spectra ({\it left}) at three
pointings, which are indicated by circles in the B-band image of the
polar-ring galaxy NGC~4650A (\citealt{whitmore87}) on the
{\it right}. \citet{watson94} estimated the mass of molecular hydrogen in
the polar ring of NGC 4650A to be $M_{\rm H_{2}} = 8-16 \times 10^{8}
\, M_{\odot}$. \copyright\ AAS. Reproduced with permission} 
\label{fig:polar_ring}
\end{center}
\end{figure}

The majority of accreting gas is perhaps in the atomic form, but
the outskirts of early-type galaxies also offer the opportunity to study
recently accreted molecular gas, which has mainly been detected in
polar rings of elliptical and S0 galaxies\index{polar-ring galaxies}
(see Fig.~\ref{fig:polar_ring} for an example). These polar rings
are present in about $0.5\%$ of nearby S0 galaxies
(\citealt{whitmore90}). CO has been detected in polar rings at 
galactocentric radii of $12 \, {\rm kpc}$ in NGC~660 (\citealt{combes92})
and $2 \, {\rm kpc}$ in NGC~2685 (\citealt{schinnerer02}; see also
\citealt{watson94, galletta97, combes13}). Published values
for the mass of molecular hydrogen in the polar rings range from $8-11
\times 10^{6} \, {M_{\odot}}$ in NGC~2685 (\citealt{schinnerer02}) to
$10^{9} \, {M_{\odot}}$ in NGC~660 (\citealt{combes92}), although the
handful of polar rings with CO detections are likely biased towards
high $M_{\rm H_{2}}$. 

Polar rings are likely caused by tidal accretion from, or a
merger\index{galaxy merger} with, a gas-rich companion and are stable on
timescales of a few Gyr as a result of self gravity
(\citealt{bournaud03}). The molecular gas observations generally support
this hypothesis because the molecular gas masses are consistent with
those of a dwarf or spiral galaxy (\citealt{watson94, galletta97, schinnerer02}).

Mergers between an early-type galaxy and a gas-rich companion can
manifest in non-polar ring systems as well. \citet{buta95} studied the
spheroid-dominated spiral galaxy NGC~7217 and concluded that most of
the molecular mass is in an outer star-forming ring at $R_{\rm Gal}
\sim 0.6 \, r_{25}$ that could have an H$_{2}$ mass that is equal to
or greater than the H{\sc i} mass. More recent work by
\citet{silchenko11} indicates that minor mergers may be responsible
for the outer ring structures.

Molecular gas has also been detected in shells at a galactocentric
radius of $15 \, {\rm kpc}$ ($1.16 \, r_{25}$) in the elliptical
galaxy Centaurus A (\citealt{charmandaris00}). \citet{charmandaris00}
calculated the mass of molecular hydrogen in the CenA shells to be
$M_{\rm H_{2}} = 4.3 \times 10^{7} \, M_{\odot}$. Like polar rings,
shells are likely caused by galaxy interactions\index{galaxy interactions} and
\citet{charmandaris00} concluded that CenA interacted with a
massive spiral galaxy rather than a low-mass dwarf galaxy because of
the large total gas mass and large ratio of molecular to atomic gas in
CenA. Additional molecular cloud formation may have been triggered
by the interaction between the shells and the CenA radio jet (see
also \citealt{salome16}).

\section{Molecular Gas Observations in Galaxy Groups and
  Clusters}\index{group environment}\index{cluster environment}\label{sec:groups}

Extended H{\sc i} gas disks beyond optical edges are common around spiral galaxies,
and as already discussed, some stimulus seems necessary to accelerate molecule formation there.
In the group/cluster environment, galaxy interactions and interactions
with the intergalactic medium (IGM)\index{intergalactic medium (IGM)}
are triggers for the H{\sc i} to H$_2$ phase transition. In the nearby M81
triplet (M82, M81, and NGC 3077), tidal interactions stretch the
atomic gas in the outskirts into tidal spiral arms, leading to
gravitational collapse to form molecular gas and stars
(\citealt{Brouillet:1992aa, Walter:2006aa}). Even an interaction with a
minor partner can be a trigger, e.g., in the M51 system, CO emission
is detected along the tidal arm/bridge between the main galaxy NGC
5194 and its companion NGC 5195 (\citealt{Koda:2009wd}). 

Interaction with the IGM in clusters is also important for the gas phase transition.
Most H{\sc i} gas in galaxy outskirts is stripped away by the ram pressure from the IGM 
(\citealt{van-Gorkom:2004aa}), while the molecular gas, which resides
mostly in inner disks, remains less affected (\citealt{Kenney:1989aa, Boselli:1997aa}).
Some compression acts on the molecular gas near the transition from
the molecular-dominant inner disks to the atomic-dominant outer disks, as
the  extents of molecular disks are smaller when the H{\sc i} in the
outskirts is stripped away (\citealt{Boselli:2014ab}). 

The stripped gas in the outskirts is seen as multiphase and has been
detected in H{\sc i} (e.g., \citealt{Chung:2009aa}), H$\alpha$ (e.g., \citealt{Yagi:2010ve}), 
and X-rays (e.g., \citealt{Wang:2004aa, Sun:2010aa}). 
Stripped molecular gas is found in NGC 4438 and NGC 4435, which are
interacting galaxies in the Virgo cluster (\citealt{Vollmer:2005aa}).
CO emission has also been discovered in the trailing tails of the
stripped gas from the disk galaxies ESO137-001 and NGC~4388 in the
Norma and Virgo clusters, respectively
(\citealt{Jachym:2014aa,verdugo15}).

The ram pressure from the IGM can also heat up and excite H$_2$
molecules, and  H$_2$ rotational emission lines\index{H$_2$ emission}
are detected  in the mid-infrared in spiral galaxies in the Virgo cluster 
(\citealt{Wong:2014aa}). The emission from warm H$_2$ is also detected
over large scales in the intergalactic space of Stephan's Quintet
galaxy group with the {\it Spitzer Space Telescope} (\citealt{Appleton:2006aa}).  
An analysis of the rotational transition ladder of its ground vibrational state
suggests the molecular gas has temperatures of $185\pm 30$\,K and
$675\pm 80$\,K. This H$_{2}$ emission coincides with and extends along
the X-ray-emitting shock front that is generated by the galaxy NGC
7318b passing through the IGM at a high velocity. 

A final example of the cluster environment affecting molecular
gas formation is that CO has been detected in cooling flows in the
outskirts of galaxies in cluster cores (e.g., \citealt{salome06}).
Clearly, the group and cluster environments produce some triggers for
the formation of molecular gas in galaxy outskirts and therefore
represent another extreme environment where we can test our
understanding of the physics of the ISM and star formation.

\section{Conclusions and Future Directions}\index{future}\label{sec:future}

Throughout the Chapter, we have highlighted that some stimuli seem
necessary to accelerate the formation of molecular gas in galaxy
outskirts. In the outskirts of the MW, stimuli include spiral arm
compression, expanding shells from supernova remnants, and
interactions with HVCs (\citealt{yasui06,izumi14,Koda:2016aa}). These same
processes are likely at play in the outskirts of
extragalactic disk galaxies. In particular, spiral density waves,
interactions, and/or cold accretion may stimulate molecule formation and the
subsequent star formation activity in XUV disks
(\citealt{thilker07,bush08,holwerda12}). Interactions and mergers likely
cause the polar rings in the outskirts of S0 galaxies, although it may
be more likely that the molecules form in the gas-rich companion
before the merger (\citealt{bournaud03}). Finally, in groups and clusters,
interactions and ram pressure stripping may accelerate molecular gas
formation in some localized areas of galaxies even as the overall
effect is to remove the star-forming fuel from the galaxies
(\citealt{Vollmer:2005aa,Jachym:2014aa}). Galaxy outskirts offer
opportunities to study the formation of molecular gas over a variety
of conditions and will be the key to understanding if there are
different modes of star formation. 

Fundamental questions remain about the physical conditions of the
ISM in the outskirts. Where is the molecular gas? What
are the basic properties of the molecular clouds, e.g., the H$_{2}$
volume density, H$_{2}$ column density, temperature, mass, and size?
How do these properties differ from the properties of molecular clouds
in the inner regions of galaxies? Is the transition from H{\sc i} to H$_{2}$
and the transition from H$_{2}$ to stars more or less efficient in the
outskirts? Are these phase transitions affected by different
large-scale processes, stimuli, or environmental conditions compared
to inner regions? Measurements of molecular gas properties often
depend on assumptions about the gas properties themselves. Right now,
those assumptions are based on our knowledge of molecular gas in
inner disks. Those assumptions need to be revisited and adjusted
continuously as we learn more about molecular gas in the
outskirts. This iterative improvement of our knowledge is now starting
in the field of galaxy outskirts. 

Building on the research that has already been done, we have
identified a number of specific studies that would begin to address
the fundamental questions above. In the outskirts of the MW, we can
study whether the relationship between the mass of the molecular cloud
and the most massive associated star is different than in the inner
MW. In the outskirts of extragalactic disk galaxies, we need to
measure the mass and size functions of molecular clouds and compare to
the MW results. In addition, theoretical studies can work towards
predicting where and how molecular gas will form in the outskirts. To
test these predictions, we encourage sensitive and wide-area mapping
of CO and/or dust continuum emission. Higher resolution (cloud-scale)
maps of H{\sc i} may also be required to accurately locate potential sites
of molecular gas formation. After each discovery of molecular gas,
subsequent multi-wavelength studies including excitation ladders of
molecular line emission are necessary to refine our knowledge of the
physical conditions of molecular gas there. In early-type galaxies, we
should search for molecular gas in XUV disks, as XUV emission could be
even more common in early-type galaxies than late-type galaxies
(\citealt{moffett12}). We hope those researchers will take note and learn
from the high failure rate of previous (published and unpublished)
searches for molecular gas in the outskirts of disk galaxies.

% Acknowledgement
% ===============
\begin{acknowledgement}
We are grateful to Fran\c coise Combes, Jennifer Donovan Meyer, Natsuko
Izumi, and Hiroyuki Nakanishi for their advice, reading suggestions,
and comments. We also thank Natsuko Izumi for allowing us to use her
figure for the distribution of molecular clouds in the outer MW
(Fig.~\ref{fig:izumi}). JK thanks the NAOJ Chile observatory, a
branch of the National Astronomical Observatory of Japan, and the
Joint ALMA Observatory for hospitality during his sabbatical visit. 
JK acknowledges the support from NASA through grant NNX14AF74G.
\end{acknowledgement}

%\bibliography{referenc}
% Standard

%\printindex

%\input{referenc}
\end{document}